# Trading Payoffs to Enlarged Neighborhoods? A New Evidence from Evolutionary Game Theory

Yandi Liu[a], Na Guo[a], Yonghui Li[b*]

a) School of Finance, Tianjin University of Finance and Economics, 25 Zhujiang Rd., Hexi Dist., Tianjin, China. 300222
b*) Department of Physics, School of Science, Tianjin University, 135 Yaguan Road, Tianjin, China. 300350

**Abstract**
Population diversity is an important aspect of Prisoner's Dilemma Game (PDG) research. However, the studies on dynamic diversity and its associated cost still need further investigation. Based on a framework comprising 2-dimensional spatial evolutionary PDG, this work examines the change in a player's neighborhood by enabling each player to pay for an upgrade of their neighborhood to switch from the von Neumann to Moore neighborhood. The upgrade cost (i.e., the cost of the advanced neighborhood) plays a vital role in cooperation promotion and serves as an entry-level to screen players. The results show that a reasonable price (entry-level) supports the cooperators' survival in an environment with high dilemma strength since it allows the formation of "normal-edge-advantage-core" clusters. On the low entry-level side, the privilege of having a larger neighborhood supports cooperation if it is accessible to all the players. On the high entry-level side, encirclements of advantage defectors appear out of the cooperative clusters. To break the encirclement and enable the expansion of the advantage clusters, the entry-level should be increased to interrupt the advantage defectors. The encirclement can be observed only in the deterministic models. Stochastic simulations are provided as robustness benchmarks.

## I. Introduction

As a fruitful branch of complex science, evolutionary game theory with simulations is commonly used to explore cooperation mechanisms in a broad spectrum of applications, from biology to social science [1–4]. Mechanisms that maintain and improve cooperation have aroused considerable attention [5]. Such mechanisms are critical in ensuring sustainable and comprehensive benefits in ecosystems and human societies, despite the egoistic nature of individuals. Egoism is a powerful characteristic that prevents cooperation and is well-captured by the Prisoner's Dilemma Game (PDG), which models the conflict between an individual's goal and the population payoff [6–10]. Various cooperation-promoting mechanisms have been proposed in the past decades, and the situations leading to resisting defections and promoting cooperation have been summarized in the literature [11]. For example, major and frequently explored classes include Nowak's five rules and the recently-discovered 6[th] reciprocity mechanism based on the dynamic utility function [12].

Ecosystems and human society are diverse. Furthermore, inhomogeneity in players' social state can facilitate cooperation, since players with a high social rank can prevail against defectors' intrusion [13]. Environmental factors (guiding whether and to what degree the player's fitness depends on the environment) have been shown to positively impact the cooperation evolution in a voluntary game [14]. An example includes diversity in a reproduction capacity (i.e., imposing reproduction restrictions upon a fraction of players), which has been proven to stimulate cooperation in small-world and scale-free networks [15]. The "Opting-out" mechanism which originated from the natural feeling of humans [16], may support the coexistence of cooperators and defectors in PDG [17] or public goods games [18].

*Yonghui Li, Corresponding author. Email: yonghui.li@tju.edu.cn

When the scope is limited to the neighbors' impact, diversity can be achieved by assigning a different number of neighbors to players [19]. The interaction range influences the cooperation evolution [20], and the neighborhood diversity can be viewed as the information advantage since the players with larger neighborhoods have the privilege to link to, interact with, and learn from more players. Furthermore, the interaction neighborhood (IN) and the learning neighborhood (LN) can differ. It was shown that cooperative clusters change with variations in the neighborhood range [21].

However, even when the neighborhood range can be optimized, it can be unaffordable to players. Maintaining or expanding a neighborhood typically consumes many resources, i.e., time or money. This participation cost may be shown as, for example, opportunity cost, membership fee, or information cost. Is it worth investing in the privilege of having more social relationships? Could the whole system benefit from such an investment? Studies identified a positive impact of participation costs as long as the benefit-to-cost (the participation cost) ratio exceeds the number of neighbors in a death-birth model [22,23]. However, a (too) high participation cost can outbalance the additional payoffs that the advantaged players could gain and, consequently, decrease the willingness to purchase the advantage [24,25]. Thus, there is a participation cost cutoff value related to achieving the optimal cooperation level. Furthermore, the participation cost determines the number of game rounds needed to achieve the cooperation equilibrium [26]. This work studies neighborhood diversity and costly participation in an imitating system.

## II. Model Definition

The standard evolutionary PDG does not consider the costs of neighborhood maintenance. The payoff matrix between two players is defined by presuming that the punishment equals zero ($P = 0$), and the sucker's payoff is -1 ($S = -1$) in each game round as follows:

$$PDG = \begin{pmatrix} R & S \\ T & P \end{pmatrix} = \begin{pmatrix} b - c & -c \\ b & 0 \end{pmatrix} = \begin{pmatrix} b - 1 & -1 \\ b & 0 \end{pmatrix}. \quad (1)$$

The "game cost" $c$ (not to be confused with the neighborhood cost) is set to 1 for simplicity. Thus, there is only one parameter ($b$) allowed to change. The dilemma strength (based on the improved definition [2,11,27,28]) measures the disturbance of the cooperation promotion and can be evaluated using two scalable indicators as follows:

$$D'_g = \frac{T-R}{R-P} = \frac{1}{b-1}; \quad D'_r = \frac{P-S}{R-P} = \frac{1}{b-1}. \quad (2)$$

Here, $D_g'$ and $D_r'$ represent the strength of the gamble-intending dilemma (GID) and risk-averting dilemma (RAD), respectively [29]. A larger Dg' (Dr') indicates that the players are eager to the temptation (averse to the sucker's payoff). With the given parameters, the PDG considered in this work lies at the balance point between GID and RAD. As benefit $b$ increases, the two strengths decrease for the fixed $S$. Hence, the donor-recipient game (DRG) [30,31], which is a subclass of the PDG, is primarily considered in this work.

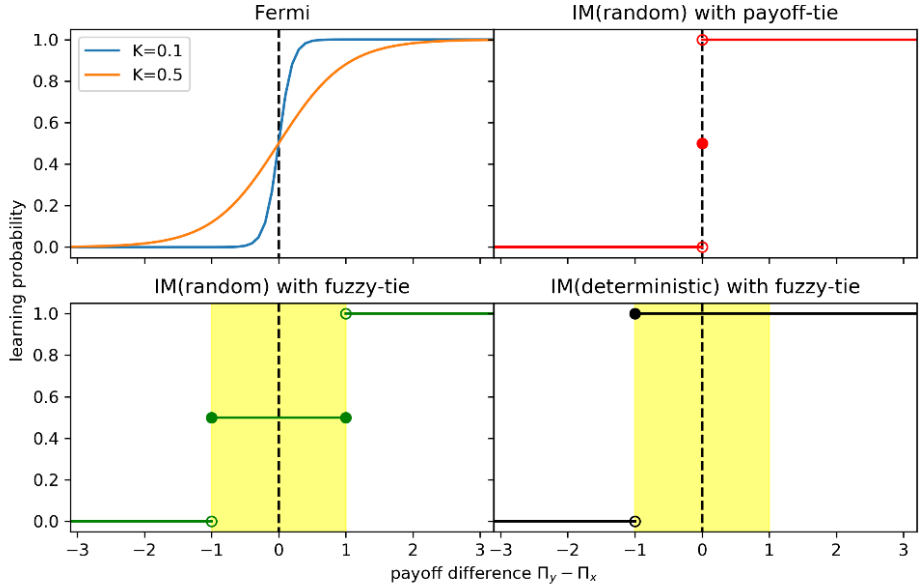

*Figure 1 Learning probability of different strategy adoption rules: (top left) Fermi rule with stochastic noise K; (top right) IM rule with random selection to resolve the payoff-tie; (bottom left) IM rule with random selection to resolve the fuzzy-tie which is seen when the payoff difference between players is 1 or less; (bottom right) IM rule with deterministic selection (e.g. choose C always) to resolve the fuzzy-tie which is seen when the payoff difference between players is 1 or less. The horizontal axis represents the payoff difference of two focal players, x and y with their payoffs $\Pi_x$ and $\Pi_y$.*

A strategy adoption rule in an evolutionary PDG model can be probabilistic, such as the Fermi rule, or deterministic, such as the "Imitation Max" (IM) rule [32,33] (also known as the "best-takes-all" rule [34] or best-response), where the decision of the best-performing neighbor is copied. Compared to the pairwise Fermi rule (or Fermi rule) in which a random player is selected with a rate of rejection, the extra characters in an IM rule are the determination (the best player is definitely picked) and resolution of a payoff-tie when different neighbors earn the same payoffs with different choices. The determination of the payoff-tie can be a "fuzzy" one [35], while the payoff-tie can be resolved by a player's conditional, unconditional, or random response (Figure 1). It would be more rational for a player with an IM rule to scan all the neighbors and for a player with the Fermi rule to check only random neighbors. In this work, the IM rule with fuzzy-tie is used, which is based on our previous work. The Fermi rule is used for consistency check.

The neighborhood size variance and the associated cost are introduced by distinguishing normal and advantage players depending on whether they utilize the von Neumann (4 neighboring players) or Moore (8 neighboring players) neighborhood. Within this work, the IN is the same as the LN and neighborhood upgrade affects both IN and LN simultaneously. The evolution begins with all the players using the von Neumann neighborhood (i.e., normal players). Furthermore, 50% of the players are randomly selected and labeled as cooperators. In each evolution round, each player can be promoted/maintained to an "advantage player" for the one proceeding round. Such a transition occurs when the player earns a payoff that surpasses the "entry-level" (*EL*), i.e., the price of the privilege of interacting with more neighbors. The *EL* price is then subtracted from the player's payoff in the next round. With the IM rule, players copy the strategy from their neighboring players with the highest payoffs (negative payoff means no impact). Consequently, the player who pays the cost may be worse off than their neighbors if the cost is too high. While this behavior may be characteristic of greedy people in reality, in the developed model, each player whose payoff is sufficiently high obtains the privilege without exceptions for simplicity. The advantage only takes effect in the next round. To possess a stable advantage, a player's payoff must be continuously and significantly growing.

As mentioned in the title, this work focuses on the competition between a higher impact and a wider neighborhood. As the evolution progresses, the payoff accumulates, while the *EL* is continuously deducted from the payoff whenever possible. For simplicity, we assume that all players follow the same rule: if the player can afford the *EL* cost of the Moore neighborhood, the player will acquire it. The rule is summarized in equation 3.

$$\begin{cases} payoff \geq EL \to \text{Moore Neighborhood} \\ payoff < EL \to \text{von Neumann Neighborhood} \end{cases} \quad (3)$$

With *EL* defined, the dynamic of pursuing the advantage can be described by the flowchart in Figure 2. Compared to a typical evolutionary PDG, the present model adds a new branch paralleled to the strategy adoption stage. This branch enables each player to obtain the advantage when the player's payoff is enough to pay the *EL* cost. The saving (unspent payoff), on the other hand, keeps the impact on the player.

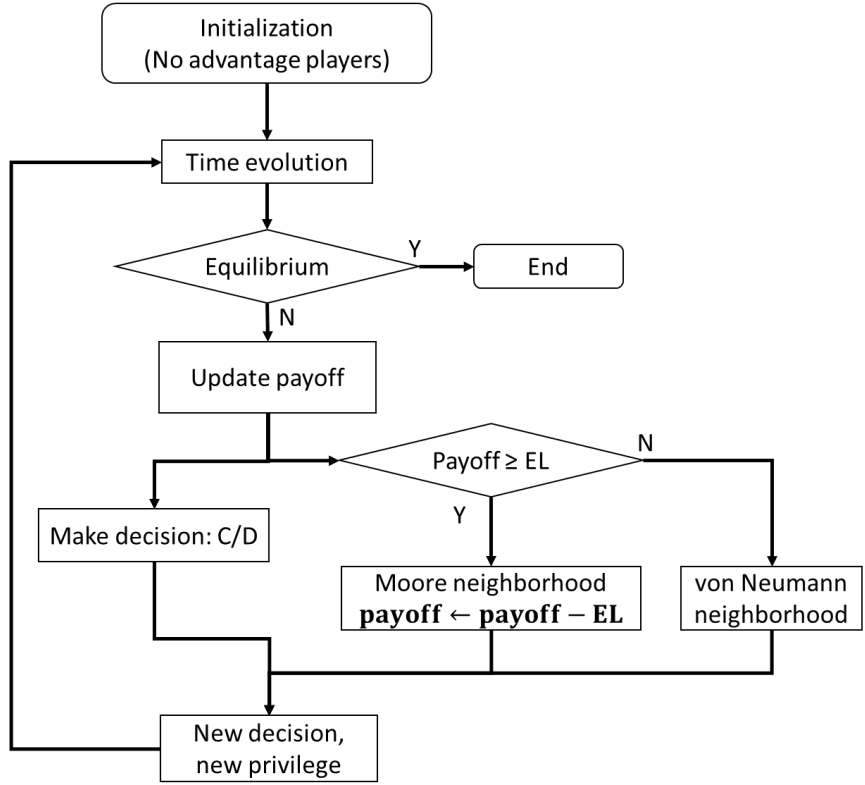

*Figure 2 Flowchart of a typical PDG evolution with the purchase of advantage.*

This work utilizes a system with $50 \times 50$ players and the IM rule for strategy updating. The evolution consists of 200-2000 Monte Carlo steps to reach equilibrium. The lattice boundary is periodic so that all players are equivalent in their local connectivity to other players. The discussion on scalability is not considered within this work. Each evaluation of the cooperation fraction ($F_c$) is obtained by averaging the values of $F_c$ over an ensemble of 50–200 independent simulations at equilibrium. In particular, the impact of EL and its interaction with the level of benefit (*b*) in promoting cooperation is studied.

## III. Numerical Simulation Results

The players convert their payoffs into privilege for the larger neighborhood as soon as they can. Such a greed for privilege interacts with the IM rule in the strategy adoption step to change the cooperation. The fraction ($F_c$) varies under the influence of *b* at different *EL* values. Figure 3 shows the evolutions of the system with different levels of benefit and *EL* values. In cases where *b* is large (i.e., when *b* equals 4.0 or 3.5), the impact of *EL* is not significant due to the dilemma strength being much lower. When the dilemma strength is much higher, the evolutions end in an equilibrium, filled with defectors when *b* is set to 2.5 and *EL* is as low as 2 without savings. In contrast, setting a much higher *EL* (i.e., *EL* = 7) improves cooperation in some cases.

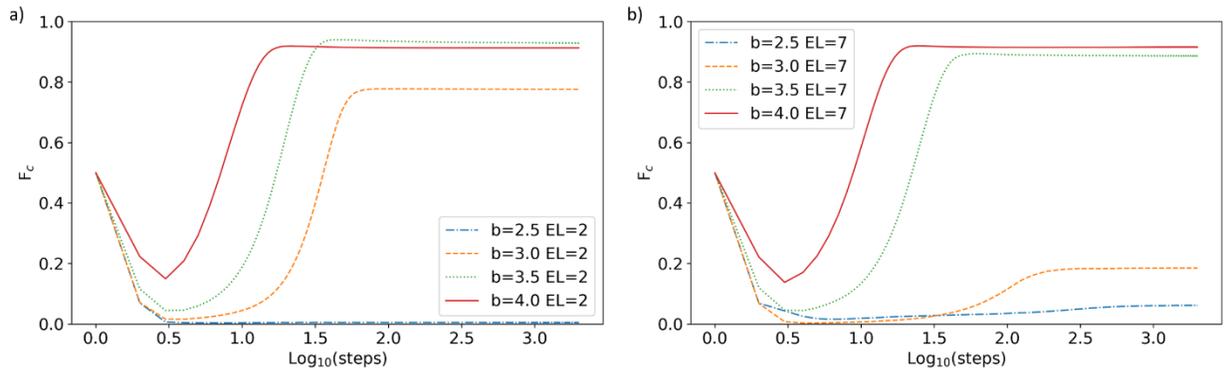

*Figure 3 Time evolutions of $F_c$ in simulations with 2500 players and different benefit (b) and entry-level (EL) values. A low EL value (EL=2) leads to no cooperation when the dilemma strength is high (b=2 or 3) but a high EL value (EL=5) improves such results.*

The evolution equilibrium depends on the value of *b*. Therefore, the next step in the study explored how $F_c$ changes with respect to b. As shown in Figure 4, the $F_c$ transitions from low to high levels as the dilemma strength decreases. When *b* is relatively high (b > 3.5, Dg' is lower than 0.4), most players are cooperators. The discontinuities in transitions emerge due to the modeled IM rule, as discussed in the literature [33]. However, since the neighborhood advantage is obtained dynamically throughout the evolution within this work, a monotonic change in $F_c$ results in some minor discontinuities.

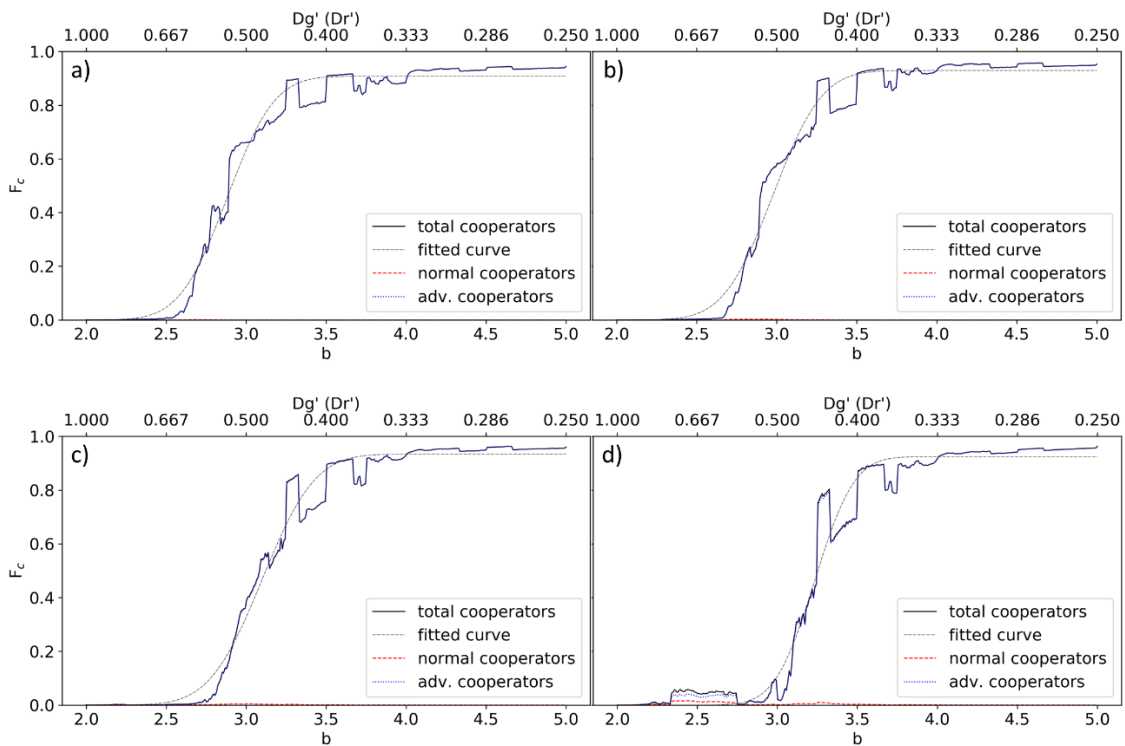

*Figure 4 The changes in $F_c$ as a function of b for different values of EL: (a) EL=1, (b) EL=3, (c) EL=5, and (d) EL=7. Minor discontinuities are observed, indicating the interaction between the IM rule and the purchase of advantage. It is rare to observe normal cooperators except in the cases with high EL and high dilemma strength (EL=7, b=2.5). Such a unique pattern is explored in the later section. Due to the setup of the model, it is rare to find normal cooperators except in (d). So in most of the subfigures, normal players vanish and all cooperators have advantage.*

Figure 4 provides additional insights into the interaction between the IM rule and the purchase of advantage. Due to the greedy attitudes of the players, the cooperators in most simulations are advantaged players. It should be noted that when *EL* is low, almost all the players can access the advantage, and normal cooperators are not present in such systems (see overlaps of dotted and solid lines in Figure 4). As the value of *EL* increases, $F_c$ decreases in the cases with high dilemma strength. Thus, in most cases, the impact of *EL* on the level of cooperation is negative. A larger *EL* significantly reduces the possibility of broadcasting cooperation for a cooperator. When $EL = 7$, the system shows the best performance in the region of high dilemma strength ($2.3 < b < 2.8$) where the population of cooperators is not dominated by normal players. Such results indicate that the normal players are important for the system to survive in high dilemma strength.

The impact of benefit (or dilemma strength) can be more intuitively evaluated by fitting the $F_c$ curve to a simpler function. Thus, this work utilizes the Cumulative Distribution Function (CDF) of the normal distribution. An additional parameter, $A_0$, is added to CDF to fit the low dilemma strength region. The modified CDF function has the following form:

$$f(x) = \frac{A_0}{2}\left[1 + erf\left(\frac{x-\mu}{\sqrt{2}\,\sigma}\right)\right], \quad (3)$$

where $erf$ denotes the error function, and $\mu$ and $\sigma$ are normal distribution parameters. The parameter $\mu$ can be interpreted as the "front" separating the low and high cooperation level zones. The parameter $\sigma$ represents the "width" of the transition zone from low- to high-level cooperation. These simplifications facilitate the model interpretation and understanding. In the following discussion, the transition zone is denoted as $b \in [\mu - \sigma, \mu + \sigma]$ or simply $\mu \pm \sigma$.

When $EL = 1$, the transition zone is $2.88 \pm 0.233$. As $EL$ increases to 3, 5, and 7, the transition zones shift to $2.97 \pm 0.24$, $3.10 \pm 0.26$, and $3.23 \pm 0.19$, respectively. By definition, it is evident that a lower front ($\mu$) and a wider width ($\sigma$) yield a more robust system to sustain higher dilemma strength. As $EL$ increases, the front values significantly decrease. However, the impact of $EL$ on the fraction of cooperation is not always negative. Thus, the interaction is further evaluated by examining the cooperation fraction values for different combinations of $EL$ and $b$. The results are shown in the color plot in Figure 5. A brighter color indicates a higher level of cooperation, while the darker regions correspond to the zones of lower cooperation.

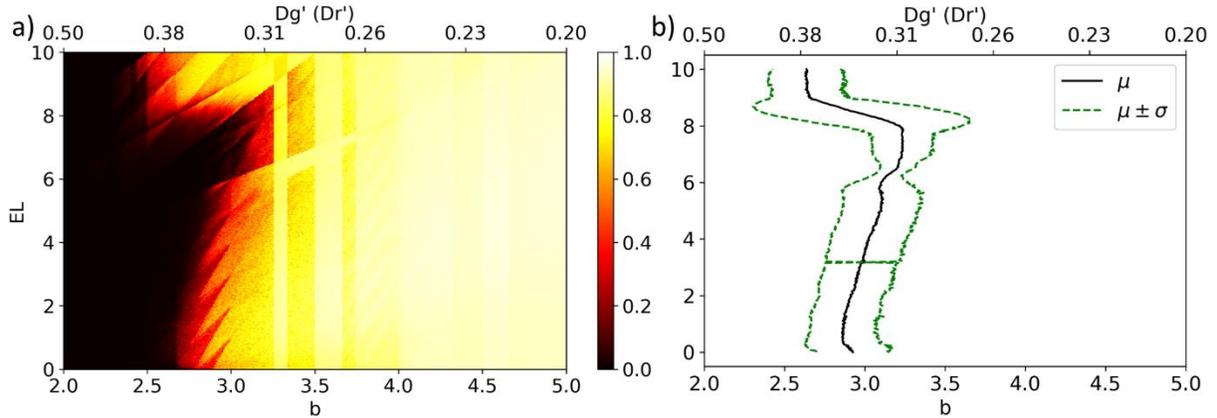

*Figure 5 (a) The color plot of the cooperation fraction for different $EL \in [0, 10]$ and $b \in [2, 5]$. Here, 360 different EL and 330 distinct b values were sampled. For each combination of EL and b, an ensemble of 200 evolutions was performed, and $F_c$ was evaluated by averaging over the ensemble. (b) The position of the front using the fitted CDF curve.*

The color plot highlights a zone of the best $F_c$ for the particular $EL$ values. When $EL$ is low, increasing its value may reduce the level of cooperation. However, when $EL$ is as high as 8, increasing the entry-level might not improve cooperation. Using the CDF fitting, such trends can be shown clearly by the front of the fitting parameter ($\mu$) and width ($\sigma$) as shown in Figure 5b.

At this point, the discussion can be divided into low $EL$ impact and high $EL$ impact. Low $EL$ impact is targeted to "fairness". Thus, as a policymaker, it is vital to restrict the $EL$ value to a specific range to maintain the high cooperation rate. Firstly, the majority should have the right to access the advantageous information but be filtered by a "fair" $EL$. Secondly, the model displays several non-desirable situations where no normal players are present in the model. $EL$ behaves as an elimination mechanism depriving the potential defective players of the advantage.

Moreover, a high *EL* impact shows a significantly different mechanism. Thus, these cases are investigated in more detail. Normal players' emergence and their impact on cooperation could be further studied by selecting two particular evolution with *EL* as high as 7 and 10. Several of the snapshots from two randomly picked evolutions are shown in Figure 6.

The figure shows the emergence of a stable structure characterized by advantaged players forming cluster cores surrounded by normal players. Such "normal-edge-advantage-core" clusters (NEACCs) have been extensively discussed in previous studies [36]. The described mechanism occurs in scenarios with a high dilemma strength. However, for such a structure to appear, a suitable *EL* value is essential. Namely, studies have shown that the NEACC does not appear in evolutions with low *EL* values (e.g., when $EL < 6$). The expandability of NEACCs is significantly affected by large *EL* values. When $EL = 7$, NEACCs do not expand for those surrounded by advantage defectors. Such confinements do not show when $EL = 10$, since a high *EL* value is not affordable for the defectors on regular basis. Therefore, a larger *EL* helps the expansion of the clusters by limiting the siege of advantage defectors.

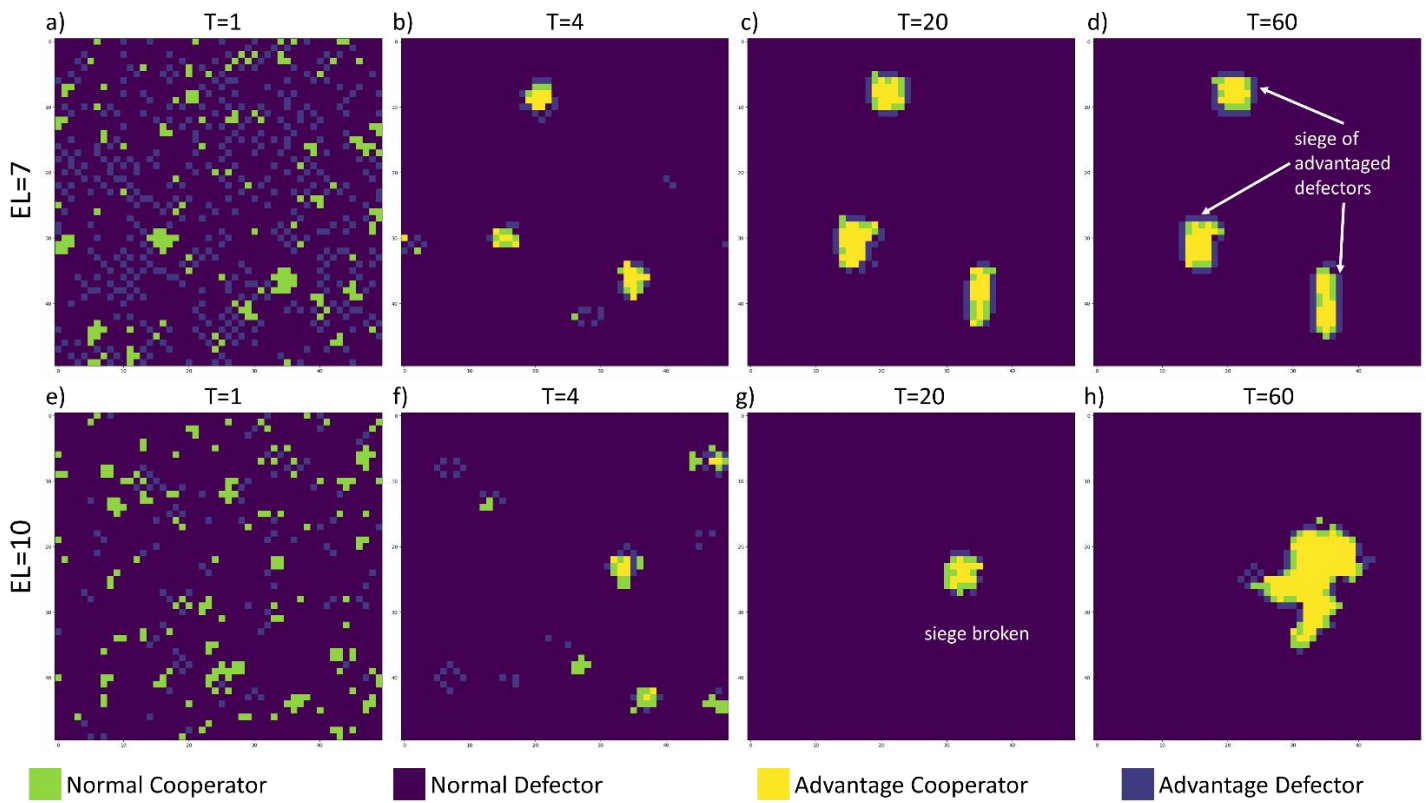

*Figure 6 Snapshots taken from randomly picked evolutions with a high dilemma strength (Dg' = 0.38 when b = 2.5). (a)-(d): snapshots at different MC steps when EL=7. (e)-(h): snapshots at different MC steps when EL=10. Unlike the low EL simulations, clusters contain advantaged players in their cores surrounded by normal players. When EL=10, the cluster expands, compared to the cases with clusters when EL=7. The expandability of the clusters depends on the advantage defectors' ability to afford the price of EL continuously.*

To verify the consistency, high *EL* impact is evaluated in similar systems. The two modified systems considered here are: A. a system with Fermi strategy updating rule (other parameters and algorithms are identical), B. a system with 12-neighbor-advantage (other parameters and algorithms are identical). The color plot of $F_c$ for b and *EL* is shown in Figure 7. Compared to Figure 5a, the results show similar trends with minor differences. In modified system A, the low $F_c$ zone expands to the high dilemma strength side, while the low $F_c$ zone shrinks to the low dilemma strength side in modified system B. The low *EL* impact is reproducible across different systems, but the high *EL* impact is not universal. The color plot also shows that the cluster expansion, enabled by *EL*, only exists in modified system B: when b = 2.5, compare with $EL = 3$, $EL = 6$ can improve the level of cooperation with the mechanism shown in Figure 6. Such a feature is not seen in modified system A. With Fermi strategy adoption, players may make stochastic decisions, which may yield a broken defective siege circle. Therefore, the impact of *EL* may not show in stochastic cases.

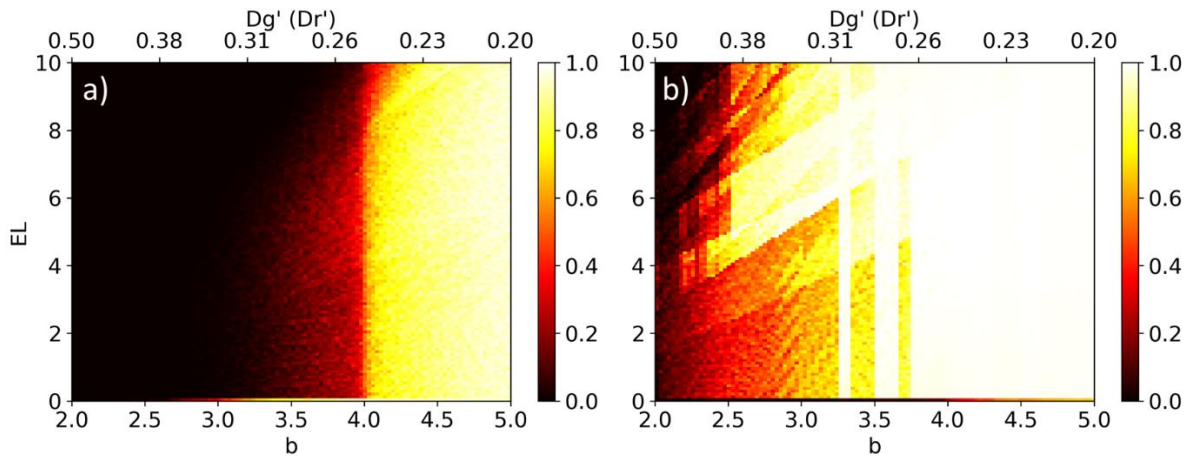

Figure 7 Consistency check. Color plots of $F_c$ for b and EL in a) modified system with Fermi strategy adoption rule and b) modified system in which advantaged players can see 12 neighbors.

## IV. Discussion and Conclusion

This work introduces neighborhood diversity by equipping players with dynamic access to neighborhood advantages in the classic 2D spatial PDG. A player that earns a sufficient payoff (> *EL*) is granted a larger neighborhood in the next round. However, the cost of the advantage is deducted from the player's payoff in the new round, lowering its influence among the neighbors (guaranteed by the Imitation Max rule). This model allows the parallel evolution of the advantage acquisition and decision making. Thus, the model enables investigation of the interaction between the *EL* values and benefit pursuit. Compared to the previous work [18], this model allows the expansion in the scopes of players, but the impacts of the players are constrained at the same time.

The non-linear impact of *EL* on the cooperation promotion was found. To develop the NEACCs, a suitable level of *EL* that represents the cost of having a large neighborhood may prevent the defective players from forming encirclements out of cooperative clusters. However, if *EL* is low, then a high dilemma strength prevents the structure formation by deleting it from the start. As a policymaker, it is important to control the *EL* value, carefully selecting the "seeds-friendly" values but finding such values is a great challenge even in the simplified model. It is also possible to set a high *EL* value when the siege of defectors is observed. A comparison between the cases with 8 and 12 neighborhoods can be reviewed as a part of the progress in the development from a local system to a well-mixed system. The fraction of the cooperator is not reduced in general. Trading payoff to a larger neighborhood is still a constraint to the players: even if the players can obtain a very large neighborhood, the impact of the player vanishes during the trade. Thus, a high level of *EL* in such trade replaces the constraint to the localization of the neighborhood.

In reality, resources are restricted and opened to certain "players." Nevertheless, it may be worth reconsidering the restriction regarding the clustering of "advantage and normal" players. These issues constitute the directions for future work. In addition, the stochastic model with probabilistic adoption and forcing the advantage purchase rule may be incompatible. It is also important to design a more reasonable stochastic rule for the advantage purchase in future investigations.

## Acknowledgments

Yonghui Li is sponsored by the National Natural Science Foundation of China (Grant No. 11804248). Na Guo has been sponsored by the National Natural Science Foundation of China (Grant No. 71903142).

**Declarations of interest: none**

**Data Availability Statement**
Data is available in the article. All of the simulated data are included in the paper as figures and/or tables.